\documentclass[a4paper,12pt]{article}
\usepackage{amsfonts}
\usepackage{amssymb,latexsym}
\makeatletter \@addtoreset{equation}{section} \makeatother

\mathchardef\varGamma="0100 \mathchardef\varDelta="0101
\mathchardef\varTheta="0102 \mathchardef\varLambda="0103
\mathchardef\varXi="0104 \mathchardef\varPi="0105
\mathchardef\varSigma="0106 \mathchardef\varUpsilon="0107
\mathchardef\varPhi="0108 \mathchardef\varPsi="0109
\mathchardef\varOmega="010A

\def\bfone{\relax{\rm 1\kern-.35em 1}}

\DeclareFontFamily{U}{rsf}{} \DeclareFontShape{U}{rsf}{m}{n}{
  <5> <6> rsfs5 <7> <8> <9> rsfs7 <10-> rsfs10}{}
\DeclareMathAlphabet\Scr{U}{rsf}{m}{n}

\newcommand{\rE}{\mathrm{E}}

\begin{document}

\begin{titlepage}

\thispagestyle{empty}

\begin{flushright}
\hfill{CERN-PH-TH/2005-221}\\
\hfill{UCLA/05/TEP/31}
\end{flushright}

\vspace{35pt}

\begin{center}{ \LARGE{\bf
Supersymmetric completion of M--theory 4D--gauge algebra\\[5mm] from twisted tori and fluxes}}
\vspace{60pt}

{\bf  R. D'Auria$^\bigstar$, S. Ferrara$^\dag$ and M.
Trigiante$^\bigstar $}

\vspace{15pt}

$^\bigstar${\it Dipartimento di Fisica, Politecnico di Torino \\
C.so Duca degli Abruzzi, 24, I-10129 Torino, and\\
Istituto Nazionale di Fisica Nucleare, \\
Sezione di Torino,
Italy}\\[1mm] {E-mail: riccardo.dauria@polito.it,  mario.trigiante@polito.it}

$^\dag$ {\it CERN, Physics Department, CH 1211 Geneva 23,
Switzerland\\ and\\ INFN, Laboratori
Nazionali di Frascati, Italy\\and\\
Department of Physics \& Astronomy, University of California, Los Angeles, CA, USA}\\[1mm] {E-mail: Sergio.Ferrara@cern.ch}

\vspace{50pt}

{ABSTRACT}
\end{center}

\medskip

 We present the supersymmetric completion of the
M--theory free differential algebra resulting from a
compactification to four dimensions on a twisted seven--torus with
4--form and 7--form fluxes turned on.  The super--curvatures are
given and the local supersymmetry transformations derived. Dual
formulations of the theory are discussed in connection with
classes of gaugings corresponding to diverse choices of vacua.
This also includes seven dimensional compactifications on more
general spaces not described by group manifolds.

\end{titlepage}

\newpage
\section{Introduction}

Recently superstring and M--theory compactifications on manifolds
admitting globally defined spinors
\cite{gck,glmw,ccdlmz,gmw,bbdp,fmt,gmmt,fkmr} (for a recent review
on string/M-theory flux--compactifications see also \cite{g}), but
with broken supersymmetry, have renewed considerable attention in
as much as they offer examples of theories with a low--energy
effective Lagrangian exhibiting spontaneously broken local
supersymmetry as well as Higgs phases of certain gauge
symmetries.\par A popular example of this class of theories
corresponds to the compactification on twisted tori with
form--fluxes turned on \cite{ss,svw,km,kstt,df,alt,h,bb}. These
models \cite{gkp,kst,fp,adfl,dh,dgftv,vz}, depending on the choice
of fluxes, can give rise to no--scale supergravities
\cite{noscale} with partially broken supersymmetry or other type
of vacua. These solutions can have either Minkowski or AdS
geometry. In the case of M--theory it turns out that the 4--form
flux on the internal space has to be trivial. The solutions with
AdS geometry are characterized by non trivial 7--form flux and
define eleven dimensional backgrounds of Freund-Rubin type
\cite{fr}  with flat directions. This has to be contrasted with
the maximal $N=8$ model when the internal space in $S^7$ and no
flat directions remain \cite{dwn}.
\par
\par Twisted tori can be described as seven dimensional group
manifolds whose isometries are part of the gauge group of the
theory \cite{ss}. This seven dimensional gauge algebra, which is
always spontaneously broken for flat groups \cite{ss}, enlarges to
a bigger symmetry when the vectors (or their dual) coming from the
3--form are also included, thus realizing a non--trivial
28--dimensional subalgebra \cite{dft1} of the maximal rigid
symmetry $\rE_{7(7)}$ \cite{cj}. In this context we show how a
suitable choice of fluxes of M--theory, different from the torus
twist, gives rise to the $N=8$ gauged ${\rm SO}(8)$ supergravity
\cite{dwn}, which is known to exhibit an $AdS$ vacuum with maximal
unbroken supersymmetry. Other choices, which still have a
7--dimensional interpretation, are also possible and some examples
are presented here.
\par
The main issue discussed in this article (sections 2 and 3) is the
supersymmetric completion of the set of curvatures and gauge
transformations which modify the free differential algebra (FDA)
of M--theory compactified to four dimensions with fluxes
\cite{daf,f,ddf,dft1,dft2,f2}. As a result we also give the spin
$3/2$ and spin $1/2$ curvatures as well as the gauge
transformation laws of the fields and the local supersymmetry
transformations. These results allow to give the full structure of
the super--FDA, with the curvatures turned on, as it is in general
the case if we consider configurations other than the vacuum,
which, in many cases, would not even exist. This is the case for
instance of massive Type IIA supergravity which does not admit a
(zero curvature) vacuum solution. \par In section 4 we discuss
different classes of gaugings, using the embedding tensor method,
introduced and developed in \cite{dwst1,dwst1.5,dwst2,dws,dwst3},
and retrieve some examples of M--theory vacua discussed in the
literature, which do not necessary fall on group manifold
compactifications.

\section{Curvatures of the supersymmetric FDA}
In this section we give the supersymmetric FDA in four dimensions
as obtained by dimensional reduction of eleven dimensional
supergravity \cite{cjs} on a twisted seven--torus with 4--form
flux turned on \cite{dft2}.\par We shall denote by
$r,s,t=0,\dots,3$ and by $a,b,c=4,\dots, 10$ the rigid four and
seven dimensional indices, and by $\mu,\,\nu=0,\dots,3$ and
$I,J,M,N=4,\dots, 10$ the curved ones respectively. The eleven
dimensional vielbein $\mathbb{V}^{\hat{a}}$ ($\hat{a}=0,\dots,
10$) is chosen of the form:
\begin{eqnarray}
\mathbb{V}^r&=&e^{\alpha\phi}\,V^r\,\,;\,\,\,\mathbb{V}^a=V^a=\phi^a_I(\sigma^I-A^I)\,,
\end{eqnarray}
where $\sigma^I$ is the basis of 1--forms on the twisted torus,
which satisfy the Maurer--Cartan equations:
\begin{eqnarray}
d\sigma^I+\frac{1}{2}\,\tau_{JK}{}^I\,\sigma^J\wedge
\sigma^K&=&0\,.
\end{eqnarray}
The vector fields $A^I_\mu$ are the Kaluza--Klein vectors.
 The internal metric
$G_{IJ}$ is given by :
\begin{eqnarray}
G_{IJ}&=&\phi^{a}_I\,\phi_{Ja}\,.
\end{eqnarray}
The ansatz for the eleven dimensional 3--form is:
\begin{eqnarray}
A^{(3)}&=&\mathcal{A}^{(3)}+B_I\wedge V^I+ A_{IJ}\wedge V^I\wedge
V^J+C_{IJK}\,V^I\wedge V^J\wedge V^K\,,
\end{eqnarray}
where $\mathcal{A}^{(3)}$ is a four--dimensional (non-propagating)
3--form, $B_{I\mu\nu}$ are seven four--dimensional antisymmetric
tensors, $A_{IJ\mu}$ are 21 vector fields and $C_{IJK}$ are 35
scalar fields.\par
 The ansatz for the eleven
dimensional gravitino $\hat{\psi}$ reads:
\begin{eqnarray}
\hat{\psi}(x,y)&=&(\Psi_A+\eta_{Aa}\,V^a)\otimes \mu^A\,,
\end{eqnarray}
where $\mu^A$ are spinors on the twisted torus. In these notations
we will define $\eta_{AI}=\phi_{I}^a\,\eta_{Aa}$.
\paragraph{Supersymmetric Free Differential Algebra}
Let us introduce the four dimensional curvatures, written as forms
on superspace and expanded in the super--vielbein basis
$\{V^r,\,\Psi_A\}$ (rheonomic parametrization \cite{cdf}):
\begin{eqnarray}
{\Scr D}\phi_I^a&=&{\Scr D}_r\phi_I^a\,V^r-i\,\phi^b_I\,\overline{\eta}_{Ab}\gamma^5\Psi_B\,(\Gamma^a)_{AB}\,,\nonumber\\
\hat{F}^I&\equiv&
dA^I+\frac{1}{2}\tau_{JK}{}^I\,A^J\wedge  A^K+\frac{i}{2}\,\overline{\Psi}_A\gamma^5\,\wedge \Psi_B\,\Gamma^a_{AB}\,\phi_a^I=\nonumber\\
&&\tilde{F}^I_{st}V^s\wedge V^t+i\,\phi^{Ia}\,\overline{{\bf \eta}}_{Aa}\,\gamma^r\,\Psi_A\wedge V_r\,,\label{fi}\\
 F^{(4)}&\equiv &d{\mathcal
A}^{(3)}-B_I\wedge F^I-g_{IJKL}\,A^I\wedge A^J\wedge A^K\wedge A^L
-\frac{1}{2}\,e^{2\alpha\phi}\,\overline{\Psi}_A\gamma^{rs}\wedge
\Psi_A\wedge V_r\wedge
V_s=\nonumber\\&=&\tilde{F}^{(4)}_{rstu}\,V^r\wedge V^s\wedge
V^t\wedge V^u\,,\nonumber\\
 H_I&\equiv& {\Scr D} B_I+2\,A_{JI}\wedge F^J+4\,g_{IJKL}\,A^J\wedge A^K\wedge A^L
+e^{\alpha\phi}\,\overline{\Psi}_A\gamma^5\,\gamma^r\,\wedge
\Psi_B\,\Gamma^a_{AB}\,\phi_{Ia}\wedge
V_r=\nonumber\\&=&\tilde{H}_{rstI} V^r\wedge V^s\wedge
V^t+e^{2\alpha\phi}\,\overline{\Psi}_A\gamma^{rs}\,{\bf \eta}_{A
I}\wedge V_r\wedge V_s\,,\nonumber\\
F^{(2)}_{IJ}&\equiv&{\Scr
D}A_{IJ}-\frac{1}{2}\,\tau_{IJ}{}^K\,B_K-3\,C_{IJK}\,F^K-6\,g_{IJKL}\,A^K\wedge
A^L- \frac{1}{2}\,\overline{\Psi}_A\wedge
\Psi_B\,\Gamma^{ab}_{AB}\,\phi_{Ia}\phi_{Jb}=\nonumber\\&=&\tilde{F}^{(2)}_{stIJ}\,V^s\wedge
V^t+ 2\,e^{\alpha\phi}\,\overline{\Psi}_A\gamma^5\,\gamma^r\,{\bf
\eta}_{B[I}\,\phi_{J]a}\,\Gamma^a_{AB}\wedge V_r\,,\nonumber\\
F^{(1)}_{IJK}&\equiv&{\Scr
D}C_{IJK}-\tau_{[IJ}{}^L\,A_{K]L}+4\,g_{IJKL}
\,A^L=\tilde{F}^{(1)}_{rIJK}V^r+
\overline{\Psi}_A\,{\bf \eta}_{B[I}\,\phi_{Ja}\phi_{K]b}\Gamma^{ab}_{AB}\,,\nonumber\\
F^{(0)}_{IJKL}&=&-g_{IJKL}-\frac{3}{2}\,\tau_{[IJ}{}^M\,C_{KL]M}-\frac{1}{2}\,\overline{{\bf
\eta}}_{A[I}\,{\bf
\eta}_{BJ}\,\phi_{Ka}\phi_{L]b}\Gamma^{ab}_{AB}\,,\label{fs}
\end{eqnarray}
where antisymmetrization in the above formulas involve only the
internal space indices $I,J,...$. In (\ref{fs}) we have denoted by
$F^I$ the following quantities:
\begin{eqnarray}
F^I&=& dA^I+\frac{1}{2}\,\tau_{JK}{}^I\,A^J\wedge A^K\,.
\end{eqnarray}\par
  The fermionic
curvatures have the following parametrization:
\begin{eqnarray}
\rho_A&=&{\mathcal D}\Psi_A+\frac{1}{4}\,\phi^I_{a}{\Scr D}
\phi_{b I}\,(\Gamma^{ab})_{AB}\wedge\Psi_B=\nonumber\\&=&
\tilde{\rho}_{rs}\,V^r\wedge
V^s+\frac{\alpha}{2}\,\partial_r\phi\,\gamma^r{}_t\,V^t\wedge
\Psi_A+\frac{i}{8}\,e^{\alpha\phi}\,\overline{\eta}_{Ca}\gamma_r\eta_{Cb}\,(\Gamma^{ab})_{AB}V^r\wedge
\Psi_B+\nonumber\\&&+i\,e^{\alpha
\phi}\,\phi^{Ia}\eta_{AI}\,\overline{\eta}_{Ba}\,\gamma_r\,\Psi_B\wedge
V^r+\frac{1}{2}\,e^{-\alpha\phi}\,\tilde{F}^I_{st}\,\phi_{Ia}\,(\Gamma^a)_{AB}\,\gamma_5\gamma^{s}V^t
\wedge\Psi_B-\nonumber\\&&-\frac{i}{2}\,(\overline{\eta}_{Ca}\gamma_r\Psi_C)\,\gamma^5\gamma^r(\Gamma^a)_{AB}\,\wedge\Psi_B
-\frac{i}{4}\,\overline{\eta}_{Ca}\gamma^5\Psi_D(\Gamma_b)_{CD}\,(\Gamma^{ab})_{AB}\wedge
\Psi_B -\nonumber\\&&
-\frac{i}{2}\overline{\Psi}_B\gamma^5\Psi_C(\Gamma^a)_{BC}\,\eta_A-\frac{i}{24}\,e^{\alpha\phi}\,\tilde{F}^{(0)}_{abcd}(\Gamma^{abcd})_{AB}\gamma^r\Psi_B\wedge
V^r+\nonumber\\&&
+\frac{i}{3}\,\tilde{F}^{(1)}_{rabc}(\Gamma^{abc})_{AB}\left(\delta^{r}_s+\frac{1}{2}\,\gamma^{r}{}_s\right)\,\gamma^5\Psi_B\wedge
V^s-\nonumber\\&&-i\,e^{-\alpha\phi}\,\tilde{F}^{(2)}_{rsab}(\Gamma^{ab})_{AB}\left(\gamma^r\delta^{s}_t+\frac{1}{4}\,\gamma^{rs}{}_t\right)\,\Psi_B\wedge
V^t+\nonumber\\&&
+i\,e^{-2\alpha\phi}\,\tilde{H}_{rsta}(\Gamma^{a})_{AB}\left(\gamma^{rs}\delta^t_u+\frac{1}{6}\,\gamma^{rst}{}_u\right)\,\gamma^5\Psi_B\wedge
V^u-\frac{i}{3}\,e^{-3\alpha\phi}\,\tilde{F}^{(4)}_{rstu}\,\gamma^{rst}\,\Psi_A\wedge
V^u\,,\nonumber\\&&\nonumber\\
 \rho_{Aa}&=&{\mathcal D}\eta_{Aa}+\frac{1}{4}\,\phi^I_{b}{\Scr D}
\phi_{c I}\,(\Gamma^{bc})_{AB}\,\eta_{Ba}=\nonumber\\&=&
\tilde{\rho}_{Aa,r}\,V^r+\frac{1}{2}\,e^{-\alpha\phi}\,\phi^I_{(a}{\Scr
D}_r \phi_{b)
I}\,(\Gamma^{b})_{AB}\gamma^5\gamma^r\Psi_B+i\,\eta_{Ab}\,(\overline{\eta}_{Ba}\gamma^5\Psi_C)\,(\Gamma^b)_{BC}+\nonumber\\&&
+\frac{i}{4}\,(\overline{\eta}_{Cb}\gamma_r\eta_{Ca})\,\gamma^5\gamma^r(\Gamma^b)_{AB}\Psi_A-
\frac{i}{2}(\overline{\eta}_{Cb}\gamma_r\Psi_C)\,\gamma^5\gamma^r(\Gamma^b)_{AB}\,\eta_{Ba}+\frac{1}{4}\,e^{-2\alpha\phi}\,\phi_{Ia}\,\tilde{F}^I_{rs}\,\gamma^{rs}\,\Psi_A-\nonumber\\&&
-\frac{i}{3}\,\tilde{F}^{(0)}_{bcde}\left[\Gamma^{bcd}\,\delta^e_a+\frac{1}{8}\,\Gamma^{bcde}{}_a\right]_{AB}\gamma^5\Psi_B-
i\,e^{-\alpha\phi}\,\tilde{F}^{(1)}_{rbcd}\left[\Gamma^{bc}\,\delta^d_a+\frac{1}{6}\,\Gamma^{bcd}{}_a\right]_{AB}\gamma^r\Psi_B-\nonumber\\&&-
i\,e^{-2\alpha\phi}\,\tilde{F}^{(2)}_{rsbc}\left[\Gamma^{b}\,\delta^c_a+\frac{1}{4}\,\Gamma^{bc}{}_a\right]_{AB}\gamma^5\gamma^{rs}\Psi_B-
\frac{i}{3}\,e^{-3\alpha\phi}\,\tilde{H}_{rstb}\left[\delta^b_a+\frac{1}{2}\,\Gamma^{b}{}_a\right]_{AB}\gamma^{rst}\Psi_B-\nonumber\\&&-
\frac{1}{24}\,e^{-4\alpha\phi}\,\tilde{F}^{(4)}_{rstu}\,\epsilon^{rstu}\Gamma_{a|AB}\Psi_B-\frac{1}{4}\,\omega_{bc,a}\,
(\Gamma^{bc})_{AB}\,\Psi_B-\frac{i}{4}\,\overline{\eta}_{Cb}\gamma^5\Psi_D(\Gamma_c)_{CD}\,(\Gamma^{bc})_{AB}\eta_{Ba}
\,.\,\nonumber\\&& \label{ffs}
\end{eqnarray}
 In (\ref{fi}), (\ref{fs}) and (\ref{ffs}) we have denoted by
$\tilde{F}$, $\tilde{H}$, $\tilde{\rho}$ and $\tilde{\rho}_A$ the
components in superspace of the curvatures along the space--time
vielbeins.  The \emph{supercovariant} field strengths originate by
projecting these components on the $dx^\mu$ basis. \par The
7--form flux \cite{d}, which we shall denote by
$\tilde{g}_{MNPQRST}=\tilde{g}\epsilon_{MNPQRST}$, transforms in
the ${\bf 1}_{+7}$ of ${\rm GL}(7,\mathbb{R})$ enters our
discussion as an intregration constant which comes about when
integrating the eleven dimensional field equation \cite{dft2}:
\begin{eqnarray}
 d (V_7\,P)&=&-\frac{1}{4}\, F^{(1)}_{IJK}\, F^{(0)}_{PQRS}\,\epsilon^{IJKPQRS}\,,\label{dP}
\end{eqnarray}
where
\begin{eqnarray}
 P&=& \frac{1}{\sqrt{-g}}\,\epsilon^{\mu_1\dots \mu_4}\, F^{(4)}_{\mu_1\dots \mu_4}\,.\label{P}
\end{eqnarray}
 The fermion fields
$\Psi_A,\,\eta_{Aa}$ yield the four dimensional gravitino $\psi_A$
and dilatino $\chi_{Aa}$ through the following combinations:
\begin{eqnarray}
\psi_A&=&e^{-\frac{1}{2}\,\alpha
\phi}\,\Psi_A+\frac{1}{2}\,e^{\frac{1}{2}\,\alpha
\phi}\,\gamma_r\gamma_5(\Gamma^a)_{AB}\eta_{Ba}\,V^r\,,\nonumber\\
\chi_{Aa}&=&e^{\frac{1}{2}\,\alpha
\phi}\,\eta_{Aa}\,.\label{transf}
\end{eqnarray}
The above redefinitions are needed in order for the resulting
kinetic term in the four dimensional Lagrangian to be diagonal.
\section{Local symmetries}
Let us consider the eleven dimensional gauge transformation:
\begin{eqnarray}
\delta\hat{A}^{(3)}&=&d\hat{\Sigma}^{(2)}\,.
\end{eqnarray}
If we introduce the parameters
${\Sigma}^{(2)},\,\Sigma_I^{(1)},\,\Sigma_{IJ}^{(0)}$ through the
following expansion:
\begin{eqnarray}
\hat{\Sigma}^{(2)}&=&{\Sigma}^{(2)}+\Sigma_I^{(1)}\wedge
V^I+\Sigma_{IJ}^{(0)}\wedge V^I\wedge V^J\,,
\end{eqnarray}
and the gauge parameter $\omega^I$ associated with the
Kaluza--Klein vectors $A^I$, the lower dimensional theory is
invariant under the following four dimensional
tensor/vector--gauge transformations \cite{df}:
\begin{eqnarray}
\delta{\mathcal A}^{(3)}&=& d{\Sigma}^{(2)}+\Sigma_I^{(1)}\wedge
F^I\,,\nonumber\\
\delta B_I&=&{\Scr D}\Sigma_I^{(1)}+2\,\Sigma_{IJ}^{(0)}\,F^J+\omega^K\,\tau_{KI}{}^N\,B_N-12\,g_{IJKL}\omega^J\,A^K\wedge A^L\,,\nonumber\\
\delta {A}_{IJ}&=&\frac{1}{2}\,\tau_{IJ}{}^K\,\Sigma_K^{(1)}+{\Scr
D}\Sigma_{IJ}^{(0)}-2\,\omega^N\,\tau_{N[I}{}^K\,A_{J]K}+12\,g_{IJKL}\omega^K\, A^L\,,\nonumber\\
\delta A^I&=&{\Scr D}\omega^I\,,\nonumber\\
 \delta
C_{IJK}&=&-\Sigma_{M[I}^{(0)}\,\tau_{JK]}{}^M+3\,\omega^L\,\tau_{L[I}{}^M\,C_{JK]M}-4\,g_{IJKL}\,\omega^L\,.\label{gauge}
\end{eqnarray}
We have denoted by ${\Scr D}$ the covariant derivative with
respect to the gauge group with parameters $\omega^I$:
\begin{eqnarray}
{\Scr D}T_{I_1\dots I_k}&\equiv&dT_{I_1\dots I_k}+(-1)^k
\,k\,A^L\,\tau_{L[I_1}{}^K\,T_{I_2\dots I_k]K}\,.
\end{eqnarray}
We shall use a different symbol ${\mathcal D}$ to denote the
covariant derivative with respect to the spin connection.\par
 Under the
gauge transformations (\ref{gauge}) the field strengths in
(\ref{fi}) and (\ref{fs}) transform covariantly. In particular
they are invariant under the transformations parametrized by the
$\Sigma$--parameters, while transform covariantly under those
parametrized by $\omega^I$. The supersymmetry variation of the
various fields are computed by contracting the correponding
curvatures by the supersymmetry parameter $\epsilon^A$ and keeping
in mind that:
\begin{eqnarray}
i_{\epsilon^A}\,\psi_B&\equiv & \psi_B(\epsilon^A)=\delta_B^A\,,
\end{eqnarray}
except for the gravitino for which the supersymmetry
transformation is computed as \cite{cdf}:
\begin{eqnarray}
\delta \psi_A&=&\mathcal{D}\epsilon_A-\frac{1}{4}\,\phi_a^I{\Scr
D}_r\phi_{Ib}(\Gamma^{ab})_{AB}\,\epsilon_B\,V^r+
i_\epsilon\rho_A\,.
\end{eqnarray}
 As an example let us compute the
variation of the Kaluza--Klein vectors keeping just the
two-fermion terms, using the parametrization of the corresponding
field strength given in (\ref{fi}) and taking into account
(\ref{transf}):
\begin{eqnarray}
\delta A^I_\mu &=& i_{\epsilon} F^I=
i\,e^{\alpha\phi}\,\phi^{Ia}\overline{\chi}_{Aa}\,\gamma_r\epsilon_A\,V^r_\mu-i\,e^{\alpha\phi}
 \overline{\epsilon}_A\gamma^5\left[\psi_{B\mu}-\frac{1}{2}\gamma_s\gamma_5(\Gamma^b)_{BC}\,\chi_{Cb}\,V_\mu^s\right]\,(\Gamma^a)_{AB}\phi_a^I\,.\nonumber\\&&
\end{eqnarray}
 Let us now consider the
local supersymmetry transformations as deduced from the rheonomic
parametrization of the four dimensional curvatures.
\begin{eqnarray}
\delta
C_{IJK}&=&\overline{\epsilon}^A(\Gamma^{ab})_{AB}\chi_{B[I}\phi_{aJ}\,\phi_{bK]}\,,\\
\delta\phi_I^a&=&-i\,\overline{\chi}_{AI}\gamma^5\epsilon_B\,(\Gamma^a)_{AB}\,,\\
\delta B_I&=&\overline{\epsilon}_A\,\gamma_{rs}
\chi_{AI}\,V^r\wedge
V^s-2\,e^{2\alpha\phi}\,\overline{\epsilon}_A\gamma^5\gamma_r\left[\psi_B-\frac{1}{2}\gamma_s\gamma_5(\Gamma^b)_{BC}\,\chi_{Cb}\,V^s\right]\,(\Gamma^a)_{AB}\phi_{Ia}\wedge
V^r+\nonumber\\&&+2i\,e^{\alpha\phi}\,A_{IJ}\wedge g^{JK}\overline{\chi}_{AK}\,\gamma_r\epsilon_A\,V^r-\nonumber\\&&-2i\,e^{\alpha\phi}\,A_{IJ}\wedge\overline{\epsilon}_A\gamma^5\left[\psi_B-\frac{1}{2}\gamma_s\gamma_5(\Gamma^b)_{BC}\,\chi_{Cb}\,V^s\right]\,(\Gamma^a)_{AB}\phi_a^J\,,\\
 \delta
 A_{IJ}&=&2\,e^{\alpha\phi}\,\overline{\epsilon}_A\gamma^5\gamma_r\,\chi_{B[I}\phi_{J]a}(\Gamma^a)_{AB}\,V^r+
 e^{\alpha\phi}\,\overline{\epsilon}_A\psi_B\,(\Gamma^{ab})_{AB}\,\phi_{Ia}\phi_{Jb}+\nonumber\\
 &&+3i\,e^{\alpha\phi}\,C_{IJK}\,g^{KM}\overline{\chi}_{AM}\,\gamma_r\epsilon_A\,V^r-\nonumber\\&&
 -3i\,e^{\alpha\phi}\,C_{IJK}\,\overline{\epsilon}_A\gamma^5\left[\psi_B-\frac{1}{2}\gamma_s\gamma_5(\Gamma^b)_{BC}\,\chi_{Cb}\,V^s\right]\,(\Gamma^a)_{AB}\phi_a^K\,,\\
 \delta A^I&=&i\,e^{\alpha\phi}\,g^{IM}\overline{\chi}_{AM}\,\gamma_r\epsilon_A\,V^r-i\,e^{\alpha\phi}
 \overline{\epsilon}_A\gamma^5\left[\psi_B-\frac{1}{2}\gamma_s\gamma_5(\Gamma^b)_{BC}\,\chi_{Cb}\,V^s\right]\,(\Gamma^a)_{AB}\phi_a^I\,,\\
 \delta {\mathcal
 A}^{(3)}&=&e^{3\alpha\phi}\,\overline{\epsilon}_A\,\gamma_{rs}\left[\psi_A-\frac{1}{2}\gamma_s\gamma_5(\Gamma^b)_{AB}\,\chi_{Bb}\,V^s\right]\wedge \,V^r\wedge
 V^s+i\,e^{\alpha\phi}\,B_I\wedge
 g^{IM}\overline{\chi}_{AM}\,\gamma_r\epsilon_A\,V^r-\nonumber\\&&-i\,e^{\alpha\phi}\,B_I\wedge
 \overline{\epsilon}_A\gamma^5\left[\psi_B-\frac{1}{2}\gamma_s\gamma_5(\Gamma^b)_{BC}\,\chi_{Cb}\,V^s\right]\,(\Gamma^a)_{AB}\phi_a^I\,.
\end{eqnarray}
In the above formulas $\phi_I{}^a$ are the metric moduli of the
internal manifold and span the coset manifold ${\rm
GL}(7,\mathbb{R})/{\rm SO}(7)$. the scalar $\phi$ is related to
the volume $V_7$ of the torus:
\begin{eqnarray}
V_7&=&{\rm det}(V_7)=e^{-2\alpha\phi}\,.
\end{eqnarray}
The value of $\alpha$ is tipically fixed to $7/3$ in order for the
bosonic fields to have the standard grading with respect to the
${\rm O}(1,1)$ rescaling of $V_7$:
\begin{eqnarray}
\phi&\rightarrow &\phi-\beta\,.\label{o11}
\end{eqnarray}
The grading of the various fields, for $\alpha=7/3$, can be
computed to be:
\begin{eqnarray}
&&A^I\,\,[-\frac{9}{7}\alpha=-3]\,\,;\,\,\,A_{IJ}\,\,[-\frac{3}{7}\alpha=-1]\,\,;\,\,\,B_I\,\,[-\frac{12}{7}\alpha=-4]\,\,;\,\,\,\phi_I^a\,\,[\frac{2}{7}\alpha=\frac{2}{3}];\,
\nonumber\\
&&V^I\,\,[-\frac{9}{7}\alpha=-3]\,\,;\,\,\,V^a\,\,[-\alpha=-\frac{3}{7}]\,\,;\,\,\,\tau_{IJ}{}^K\,\,[\frac{9}{7}\alpha=+3]\,\,;\,\,\,g_{IJKL}\,\,[\frac{15}{7}\alpha=+5];\,
\nonumber\\&&\Psi_A\,\,[-\frac{1}{2}\alpha=-\frac{7}{6}]\,\,;\,\,\,\eta_{Aa}\,\,[\frac{1}{2}\alpha=\frac{7}{6}]\,\,;\,\,\,\psi_A\,\,[0]\,\,;\,\,\,\chi_{Aa}\,\,[0]\,;\nonumber\\&&e^{k\alpha\phi}\,\,[-k\,\alpha=-\frac{7}{3}\,k]\,.
\end{eqnarray}
 As far as the spinor fields are concerned, their supersymmetry
transformation rules, up to three fermion terms, are:
\begin{eqnarray}
\delta\psi_A&=&{\Scr D}\epsilon_A+\frac{1}{4}\,\phi^I_{a}{\Scr
D}_r \phi_{b
I}\,(\Gamma^{ab})_{AB}\epsilon_B\,V^r-\frac{1}{8}\,e^{-\alpha\phi}\,F^I_{st}\,\phi_{Ia}\,(\Gamma^a)_{AB}\,\gamma_5\gamma^{st}\gamma_r
V^r\epsilon_B-\nonumber\\
&&+\frac{i}{16}\,e^{\alpha\phi}\,F^{(0)}_{abcd}\,(\Gamma^{abcd})_{AB}\gamma_r\,V^r\epsilon_B-
\frac{i}{2}\,F^{(1)}_{abcr}\,(\Gamma^{abc})_{AB}\gamma^5\,V^r\,\epsilon_B+\nonumber\\&&
-\frac{3i}{8}\,e^{-\alpha\phi}\,F^{(2)}_{abrs}\,(\Gamma^{ab})_{AB}\gamma^{rs}{}_t\,V^t\epsilon_B+\frac{i}{4}\,e^{-\alpha\phi}\,F^{(2)}_{abrs}\,(\Gamma^{ab})_{AB}\gamma^{r}\,V^s\epsilon_B-\nonumber\\&&
-\frac{i}{2}\,e^{-2\alpha\phi}\,H_{arst}\,(\Gamma^{a})_{AB}\gamma^5\gamma^{rst}{}_u\,V^u\,\epsilon_B
+\frac{i}{4}\,e^{-3\alpha\phi}\,F^{(4)}_{rstu}\,\gamma^{rst}\,V^u\epsilon_A-\nonumber\\
&&-\frac{1}{8}\,e^{\alpha\phi}\,\omega_{ab,c}\,(\Gamma^c\Gamma^{ab})_{AB}\,\gamma_5\gamma_r\,V^r\,\epsilon_B\,.\nonumber\\
\delta\chi_{Aa}&=&\frac{1}{2}\,\phi^I_{(a}{\Scr D}_r \phi_{b)
I}\,(\Gamma^{b})_{AB}\gamma^5\gamma^r\epsilon_B+\frac{1}{4}\,e^{-\alpha\phi}\,F^I_{st}\,\phi_{Ia}\,\gamma^{st}\epsilon_A-\nonumber\\&&
-\frac{i}{3}\,e^{\alpha\phi}\,F^{(0)}_{bcde}\left[\Gamma^{bcd}\,\delta^e_a+\frac{1}{8}\,\Gamma^{bcde}{}_a\right]_{AB}\gamma^5\epsilon_B+
i\,F^{(1)}_{bcdr}\left[\Gamma^{bc}\,\delta^d_a+\frac{1}{6}\,\Gamma^{bcd}{}_a\right]_{AB}\gamma^r\epsilon_B-\nonumber\\&&-
i\,e^{-\alpha\phi}\,F^{(2)}_{bcrs}\left[\Gamma^{b}\,\delta^c_a+\frac{1}{4}\,\Gamma^{bc}{}_a\right]_{AB}\gamma^5\gamma^{rs}\epsilon_B+
\frac{i}{3}\,e^{-2\alpha\phi}\,H_{brst}\left[\delta^b_a+\frac{1}{2}\,\Gamma^{b}{}_a\right]_{AB}\gamma^{rst}\epsilon_B-\nonumber\\&&-
\frac{1}{24}\,e^{-3\alpha\phi}\,F^{(4)}_{rstu}\,\epsilon^{rstu}\Gamma_{a|AB}\epsilon_B-\nonumber\\
&&-\frac{1}{4}\,e^{\alpha\phi}\,\omega_{bc,a}\,(\Gamma^{bc})_{AB}\,\epsilon_B\,,\label{transff}
\end{eqnarray}
where we have defined:
\begin{eqnarray}
\omega_{ab,c}&=&d_{a,bc}+d_{b,ca}-d_{c,ab}\,\nonumber\\
d_{a,bc}&=&\frac{1}{2}\,\tau_{bc,a}+\frac{i}{2}\,e^{-\alpha\phi}\,\overline{\chi}_{Ab}\,\gamma^5\chi_{Bc}\,\Gamma_{a|AB}\,\,\,\,;\,\,\,\,\,\tau_{bc,a}=\phi_b^I\phi_c^J\phi_{Ka}\,
\tau_{IJ}{}^K\,.
\end{eqnarray}
In the transformation rules (\ref{transff}) we used  the
components of the ordinary field strengths $F,\,H$ in place of the
components of the corresponding supercovariant field strengths
$\tilde{F},\,\tilde{H}$ since the final expressions would differ
by three fermion terms.
\section{The embedding tensor description applied to different classes of gaugings}
In standard four dimensional maximal supergravity the electric
$e_\Lambda$ and magnetic $m^\Lambda$ charges, where
$\Lambda=1,\dots, 28$, transform together in the ${\bf 56}$ of
$\rE_{7(7)}$ and the most general gauging can be described in
terms of an embedding tensor \cite{dwst1,dwst1.5,dwst2,dws,dwst3}
$\theta_n{}^\sigma\equiv\{\theta_\Lambda{}^\sigma,\,\theta^{\Lambda\sigma}\}$
($n=1,\dots, 56$ and $\sigma=1,\dots, 133$), which expresses the
generators $X_n$ of the gauge algebra ${\frak g}$ in terms of
$\rE_{7(7)}$ generators $t_\sigma$:
\begin{eqnarray}
X_n&=&\theta_n{}^\sigma\,t_\sigma\,.
\end{eqnarray}
In this notation consistency of the gauging requires the rank of
$\theta$, as a $56\times 133$ matrix, not to be greater than $28$
since no more that 28 gauge vectors can take part to the minimal
couplings. Supersymmetry and closure of the gauge algebra inside
$\rE_{7(7)}$ require a linear and a quadratic condition in
$\theta$ respectively \cite{dwst1,dwst1.5}:
\begin{eqnarray}
\theta &\in& {\bf 912}\,\subset {\bf 56}\times {\bf 133}\,,\label{912}\\
&&\theta^{\Lambda[\sigma}\,\theta_\Lambda{}^{\gamma]}=0\,.\label{2c}
\end{eqnarray}
The above constraints are all $\rE_{7(7)}$ covariant. The last
condition ensures that \emph{there always exists a symplectic
rotation acting on the index $\Lambda$ (electric and magnetic) as
a consequence of which all the vectors associated with the
generators $X_n$ are electric (or all magnetic).}\par Once
$\theta$ is fixed, as a solution of (\ref{912}) and (\ref{2c}),
the structure of the gauge algebra is also fixed. Indeed if we
introduce the following $\rE_{7(7)}$--tensor:
\begin{eqnarray}
X_{mn}{}^p&=& \theta_m{}^\sigma\,(t_\sigma)_n{}^p\,,
\end{eqnarray}
the gauge algebra has the following structure:
\begin{eqnarray}
[X_m,\,X_n]&=&-X_{mn}{}^p\,X_p\,.
\end{eqnarray}
In terms of $X_{mn}{}^p$, the ${\bf 912}$ in the decomposition of
${\bf 56}\times {\bf 133}$ is singled out by requiring the
constraint $X_{(mnp)}=0$ (the indices in the {\bf 56} are raised
and lowered by using the symplectic invariant matrix), which can
be taken as an equivalent formulation of condition
(\ref{912}).\par In the standard formulation of gauged
supergravity, the definition of the gauged Lagrangian is always
referred to the symplectic frame (the \emph{electric frame}) in
which the components of $\theta$ are all electric (namely in which
$\theta^{\Lambda\sigma}=0$) so that only the electric vector
fields $A^\Lambda_\mu$ are involved in the gauging. Given a
generic solution
$\theta^{\Lambda\sigma},\,\theta_\Lambda{}^{\sigma}$ of eqs.
(\ref{912}) and (\ref{2c}), the electric frame is reached by means
of a symplectic rotation whose existence, as previously stressed,
is guaranteed by eq. (\ref{2c}), and which maps:
\begin{eqnarray}
\theta_n{}^\sigma\equiv\{\theta_\Lambda{}^\sigma,\,\theta^{\Lambda\sigma}\}&\rightarrow
&\theta^\prime_n{}^\sigma\equiv\{\theta^\prime_\Lambda{}^\sigma,\,0\}\,.
\end{eqnarray}
In this frame the gauge connection will have the form
$\Omega_\mu=A^\Lambda_\mu\,
\theta^\prime_\Lambda{}^\sigma\,t_\sigma$. In a recent work
\cite{dwst3} a novel formulation of gauged supergravity was
proposed in which the Lagrangian can be written in a generic
symplectic frame, as function of both electric
($\theta_\Lambda{}^{\sigma}$) and magnetic
($\theta^{\Lambda\sigma}$) charges, coupled in a
symplectic--invariant way to (non-abelian) electric
($A^\Lambda_\mu$) and magnetic $A_{\mu\Lambda}$ gauge fields. The
new gauge connection now reads:
\begin{eqnarray}
\Omega_\mu&=&A^\Lambda_\mu\,X_\Lambda+A_{\mu\Lambda}\,X^\Lambda\,.
\end{eqnarray}
This formulation requires the introduction of 133 tensor fields
$B_{ \mu\nu\sigma}$ in the adjoint of $\rE_{7(7)}$ which enter the
Lagrangian only in the combination with the magnetic charges:
$\theta^{\Lambda\sigma}\,B_{\mu\nu\sigma}$. The advantage of this
formulation is that the $\rE_{7(7)}$--covariance of the field
equations and Bianchi identities is still manifest (provided
$\theta$ is transformad together with all the other fields). The
new fields $B_{\mu\nu\sigma}$ and $A_{\mu\Lambda}$ are described
in such a way as not to introduce any new propagating degree of
freedom. This is reflected by the fact that the corresponding
field equations (namely the equations obtained by varying the
Lagrangian with respect to the two type of fields) are
non--dynamical. In addition to this we have vector and
tensor--gauge invariance of the Lagrangian, which allow us, by
performing various kind of gauge fixing, to distribute the 128
propagating bosonic degrees of freedom among all the bosonic
fields of the theory. We refer the reader to \cite{dwst3} for the
explicit form of the field equations and the gauge transformation
laws.\par A particular case in which the non--dynamical field
equations are easily solvable is the case in which the magnetic
components of the embedding tensor contract only isometries $t_i$
of the scalar manifold which act as translations on a set of
corresponding scalar fields $\varphi^i$:
\begin{eqnarray}
t_i&:&\,\,\,\,\,\varphi^i\rightarrow \varphi^i+c^i\,.
\end{eqnarray}
The index $\Lambda$ naturally splits in the couple of indices
$\Lambda=\mathcal{I}, U$, so that $\theta^{\mathcal{I}i}$ is a
non--singular square matrix. In this basis we have:
\begin{eqnarray}
\theta^{\mathcal{I}\sigma}&=&0\,,\,\,\sigma\neq
i\,\,\,;\,\,\,\,\theta^{U\,\sigma}=0\,,\,\,\forall\,\sigma\,,\nonumber\\
\theta^{\mathcal{I}[i}\,\theta_{\mathcal{I}}{}^{j]}
&=&0\,,\label{exm0}
\end{eqnarray}
and therefore only the tensors $B_{\mu\nu i}$ enter the
Lagrangian. Let us consider two ways of gauge fixing the
vector/tensor gauge invariance which are relevant to our analysis.
We can use the gauge invariance of $A_{\mathcal{I}\mu}$ to
eliminate the scalar fields $\varphi^i$, recalling that the
covariant derivative on these scalar fields read:
\begin{eqnarray}
D_\mu\varphi^i&=&\partial_\mu \varphi^i+\theta^{\mathcal{I}
i}\,A_{\mu\mathcal{I}}+\dots\,,
\end{eqnarray}
where the ellipses refer to the electric minimal coupings which
are not relevant to our discussion. Then we use one of the
non--dynamical equations of motion to eliminate
$A_{\mu\mathcal{I}}$ in favor of $B_{\mu\nu i}$. The resulting
gauge--fixed action will describe the tensors $B_{\mu\nu i}$
instead of the scalars $\varphi^i$ and the electric vectors
$A^\Lambda_\mu=\{A^{\mathcal{I}}_\mu,\,A^U_\mu\}$. In these models
the original second order constraint (\ref{2c}) becomes eq.
(\ref{exm0}) which has the form of the $e\times m=0$ constraint
found in the literature when dealing with supergravity theories
coupled to antisymmetric tensor fields
\cite{dafe,dsv,sv,s,dftv}.\par On the other hand we could start by
fixing the tensor--gauge invariance by eliminating the electric
fields $A^{\mathcal{I}}_\mu$ through their coupling terms with the
tensors $B_{\mu\nu i}$:
\begin{eqnarray}
F^\mathcal{I}_{\mu\nu}+\frac{1}{2}\,\theta^{\mathcal{I}
i}\,B_{\mu\nu i}\,.
\end{eqnarray}
Then we can use one of the non--dynamical field equations to
eliminate $B_{\mu\nu i}$ in favor of the remaining gauge fields.
The resulting theory will describe $70$ scalar fields, no
antisymmetric tensor field and the new electric vectors
$A_{\mu\mathcal{I}},\,A^U_\mu$. This procedure has thus
automatically produced the symplectic transformation which brought
our original $\theta$ to the electric frame.\par
 \paragraph{First example: M--theory compactification on twisted tori with flux}
 If we are considering toroidal compactifications of eleven
dimensional supergravity to four dimensions, the higher
dimensional origin of the four dimensional fields is specified by
branching the relevant $\rE_{7(7)}$--representations with respect
to the ${\rm GL}(7,\mathbb{R})$ subgroup associated with the
metric moduli of the seven--torus:
 \begin{eqnarray} {\bf 56}&\rightarrow &
\overline{{\bf 7}}_{-3}+{\bf
21}_{-1}+\overline{{\bf 21}}_{+1}+{\bf 7}_{+3}\,,\label{56}\\
{\bf 133}&\rightarrow & {\bf 7}_{-4}+ \overline{{\bf
7}}_{+4}+\overline{{\bf 35}}_{-2}+{\bf 35}_{+2}+{\bf 48}_0+{\bf
1}_0\,,\label{133}\\
{\bf 912}&\rightarrow &{\bf 1}_{-7}+{\bf 1}_{+7}+{\bf
35}_{-5}+\overline{{\bf 35}}_{+5}+ (\overline{{\bf
140}}+\overline{{\bf 7}})_{-3}+ ({\bf 140}+{\bf 7})_{+3}+{\bf
21}_{-1}+\overline{{\bf 21}}_{+1}+\nonumber\\&&{\bf
28}_{-1}+\overline{{\bf 28}}_{+1}+{\bf 224}_{-1}+\overline{{\bf
224}}_{+1}\,.\label{912b}
\end{eqnarray}
In the branching of the ${\bf 56}$ the $\overline{{\bf 7}}_{-3}$
and ${\bf 21}_{-1}$ define $A^I_\mu,\,A_{\mu IJ}$ respectively
while ${\bf 7}_{+3}$ and $\overline{{\bf 21}}_{+1}$ their magnetic
duals. In the branching of the adjoint representation of
$\rE_{7(7)}$
 we denote by $t_M{}^N,\,t^{MNP},\,t_P$ the generators in the ${\bf 48}_0+{\bf 1}_0,\,{\bf 35}_{+2}$
 and $\overline{{\bf
7}}_{+4}$ respectively (with an abuse of notation we characterize
each generator by the representation of the corresponding
parameter, this allows a simpler interpretation of the table
below). In the solvable Lie algebra representation of the scalar
manifold, the metric moduli $\phi_I^a$ parametrize the generators
$t_{I}{}^J$ with $I\ge J$, while the scalars $C_{IJK}$ and
$\tilde{\phi}^I$ (dual to $B_{\mu\nu I}$) are parameters of the
generators $t^{MNP}$ and $t_M$ repsectively. The ${\bf 912}$ is
the representation of the embeddig matrix, which encodes all
possible deformations (minimal couplings, mass terms) of the
ungauged $N=8$ theory. It is natural therefore to identify the
background quantities $\tau_{IJ}{}^K,\,g_{IJKL},\,\tilde{g}$ with
components of the embedding tensor. Indeed each component
representation on the right hand side of (\ref{912b}) defines a
consistent gauged supergravity. Some of them have an immediate
interpretation in terms of background fluxes, like $\overline{{\bf
35}}_{+5}$ which represents the 4--form flux $g_{IJKL}$, or
parameters related to the geometry of the internal space, like the
${\bf 140}_{+3}$, which defines the twist--tensor $\tau_{MN}{}^P$.
The structure of the gauge algebra implied by each of these
representations is best understood from the following table:
\begin{center}
{\small \begin{tabular}{|c||c|c|c|c|}\hline
    & ${\bf 7}_{+3}$ & $\overline{{\bf 21}}_{+1}$ & ${\bf
21}_{-1}$ & $ \overline{{\bf 7}}_{-3}$  \\\hline\hline
  $\overline{{\bf 7}}_{+4}$ & ${\bf 1}$ & $\overline{{\bf 35}}$ & ${\bf 140}+{\bf 7}$ &
   $\overline{{\bf 28}}+\overline{{\bf 21}}$ \\
  ${\bf 35}_{+2}$ & $\overline{{\bf 35}}$ &${\bf 140}$ & $\overline{{\bf 21}}+\overline{{\bf 224}}$ &
  ${{\bf 21}}+{{\bf 224}}$ \\
  ${\bf 48}_0$ & ${\bf 140}+{\bf 7}$ & $\overline{{\bf 21}}+\overline{{\bf 28}}+\overline{{\bf 224}}$ & ${{\bf 21}}+{{\bf 28}}+{{\bf 224}}$ & $\overline{{\bf 140}}+\overline{{\bf 7}}$\\
  ${\bf 1}_0$ & ${\bf 7}$ & $\overline{{\bf 21}}$ & ${\bf
21}$ & $ \overline{{\bf 7}}$ \\
  $\overline{{\bf 35}}_{-2}$ & $\overline{{\bf 21}}+\overline{{\bf 224}}$ & ${{\bf 21}}+{{\bf 224}}$ &
  $\overline{{\bf 140}}$ & ${\bf
35}$ \\
  ${\bf 7}_{-4}$ & ${{\bf 28}}+{{\bf 21}}$ & $\overline{{\bf 140}}+\overline{{\bf 7}}$ & ${\bf
35}$ & ${\bf 1}$ \\ \hline
\end{tabular}}
\label{t1}
\end{center}
The first row and column contain the representations in the
branchings of ${\bf 56}$ and the ${\bf 133}$ respectively, while
the bulk contains representations in the branching of ${\bf 912}$.
The table specifies the origin of the latter representations in
the branching of the product ${\bf 56}\times{\bf 133}$ and  it
should be read as ``first row times first column gives bulk''. The
grading of each entry of the table has been suppressed for the
sake of simplicity, since it coincides with the sum of the
gradings of the corresponding elements in the first row and
column.\par The gauged supergravity models describing the class of
compactifications we are considering are thus obtained by
restricting $\theta$ to the representations ${\bf
140}_{+3}+\overline{{\bf 35}}_{+5}+{\bf 1}_{+7}$. If we define on
the ${\bf 56}$ representation the following symplectic product:
\begin{eqnarray}
V^n\,W^m\,\mathbb{C}_{nm}&=&V^M\,W_M-V_M\,W^M+V_{NM}\,W^{NM}-V^{NM}\,W_{NM}\,,
\end{eqnarray}
the relevant components of the tensor $X_{mn}{}^p$ read:
\begin{eqnarray}
X_{MN}{}^{PQ}{}_{R}&=&X_{MN,R}{}^{PQ}=-\tau_{MN}{}^{[P}\delta^{Q]}_R\,\nonumber\\
X^{MN,PQ}{}_{R}&=&X^{MN}{}_R{}^{PQ}=-\frac{1}{2}\,\delta_R^{[P}\,\epsilon^{Q]MNN_1
N_2 N_3 N_4}\,g_{N_1 N_2 N_3 N_4}\,,\nonumber\\
X^{MN}{}_{PQ,R}&=&-X^{MN}{}_{R,PQ}=3\,\tau_{[PQ}{}^{[M}\delta_{R]}^{N]}\,,\nonumber\\
X^{MN,RS,LT}&=&X^{MN,LT,RS}-3\,\tau_{PQ}{}^{[M}\,\epsilon^{N]PQRSLT}\,,\nonumber\\
X_{M}{}^L{}_S&=&-X_{M,S}{}^L=\tau_{MS}{}^L\,,\nonumber\\
X_{M}{}^{RS}{}_{PQ}&=&-X_{M,PQ}{}^{RS}=2\,\tau_{M[P}{}^{[R}\delta^{S]}_{Q]}\,,\nonumber\\
X_{M,PQ,R}&=&-X_{M,R,PQ}=g_{MPQR}\,,\nonumber\\
X_M{}^{PQ,RS}&=&X_M{}^{RS,PQ}=-g_{MM_1 M_2 M_3}\,\epsilon^{M_1 M_2
M_3 PQRS }\,,\nonumber\\
X_M{}^{PQ}{}_N&=&X_{M,N}{}^{PQ}=-\tilde{g}\,\delta_{MN}^{PQ}\,.
\end{eqnarray}
One can verify that $X_{(mnp)}=0$, consistently with the ${\bf
912} $ condition. If we apply to this model the construction
outlined above, we can write a consistent theory (with manifest
$\rE_{7(7)}$ global on--shell covariance) which describes among
the other fields the tensors $B_{\mu\nu M}$ and their dual scalar
fields $\tilde{\phi}^M$ at the same time. Since in this case the
magnetic components of $\theta$ are just $\theta_{MN}{}^P=-(3/2)
\,\tau_{MN}{}^P$, if we denote by $r$ the rank of this $21\times 7
$ matrix, the only tensor fields entering the Lagrangian will be
  $r$ out of the tensors $B_{\mu\nu P}$. The gauge connection has
  the form:
  \begin{eqnarray}
\Omega_\mu&=&\tilde{A}_\mu^{MN}\,X_{MN}+A_{\mu
MN}\,X^{MN}+A^M_\mu\,X_M\,,
  \end{eqnarray}
and involves also the magnetic vector fields $\tilde{A}_\mu^{MN}$.
To obtain the model describing the tensor fields $B_{\mu\nu I}$ in
place of their dual scalar fields, namely the model discussed in
the first sections of the present paper, we can fix the gauge
invariance associated with $\tilde{A}_\mu^{MN}$ to eliminate
$\tilde{\phi}^M$, through the magnetic minimal coupling:
\begin{eqnarray}
D_\mu\tilde{\phi}^M&=&\partial_\mu
\tilde{\phi}^M+\tau_{PQ}{}^M\,\tilde{A}^{PQ}_\mu+\dots\,,
\end{eqnarray}
where the ellipses denote the electric minimal couplings. Then we
can use one of the non--dynamical field equations, which reads:
\begin{eqnarray}
\tau_{PQ}{}^M\,\epsilon^{\mu\nu\rho\sigma}\,(\partial_\nu
B_{\rho\sigma M}+\dots) &\propto &
G_{RS}\,\tau_{PQ}{}^R\,(\tau_{NL}{}^S\,\tilde{A}^{NL\mu}+\dots)\,,\label{nondyn}
\end{eqnarray}
to eliminate $\tilde{A}^{MN}_\mu$ in favor of $B_{\rho\sigma M}$
in the Lagrangian. The ellipses on the left hand side of
(\ref{nondyn}) denote topological terms involving the electric
vector fields, while the ellipses on the right hand side stand for
electric minimal couplings.\par We could have proceeded
differently by first fixing the tensor gauge invariance associated
with $B_{\mu\nu M}$ in order to eliminate $r$ of the $A_{\mu MN}$
fields and then expressing the tensor fields in terms of the
remaining vector fields through one of the non--dynamical field
equations. This procedure would have yield  the gauged $N=8$
supergravity with no tensor field and 70 scalar fields considered
in \cite{dft1}. This clarifies the relation between the dual
descriptions of M-theory compactification on twisted tori with
fluxes studied in the literature, namely the model with tensor
fields in which the local symmetries are encoded in a
(supersymmetric) free differential algebra, and the dual model
without tensor fields in which the local symmetries of the
Lagrangian is described by an ordinary Lie algebra.
\paragraph{Second
example: the ${\rm CSO}(p,q,r)$ gauging as an M--theory
compactification.}
 From the branching of the ${\bf 912}$ with respect to ${\rm GL}(7,\,\mathbb{R})$, we may consider the gauged model arising from the components:
\begin{eqnarray}
\theta_{MN}=\theta_{(MN)}&\in & {\bf
28}_{-1}\,\,\,;\,\,\,\,\tau_M\in {\bf 7}_{+3}
\,\,\,;\,\,\,\tilde{g}\in {\bf 1}_{+7}\,,\label{comp}
\end{eqnarray}
$\tilde{g}$ being the 7--form flux. These entries of the embedding
tensor can be re-arranged in a single $8\times 8$ symmetric tensor
$\theta_{AB}$ ($A,B=1,\dots, 8=1,M$) transforming in the ${\bf
36}$ of ${\rm SL}(8,\mathbb{R})$, maximal subgroup of ${\rm
E}_{7(7)}$:
\begin{eqnarray}
\theta_{AB}&=&\left(\matrix{\tilde{g}&\tau_N\cr\tau_M&\theta_{MN}}\right)\,.
\end{eqnarray}
 Indeed with respect to ${\rm GL}(7,\mathbb{R})$ the
following branching holds:
\begin{eqnarray}
{\bf 36}&\rightarrow & {\bf 28}_{-1}+{\bf 7}_{+3}+{\bf 1}_{+7}\,.
\end{eqnarray}
The corresponding gauge algebra generators have the form
$X_{AB}=\{X_I,\,X_{IJ}\}$ and are gauged by the vectors
$A^{AB}_\mu=\{A^I_\mu,\,\tilde{A}^{IJ}_\mu\}$ in the  ${\bf
28}^\prime$ of ${\rm SL}(8,\mathbb{R})$. They close the following
algebra:
\begin{eqnarray}
\left[X_{AB},X_{CD}\right]&=&f_{AB,CD}{}^{EF}\,X_{EF}\,\\
f_{AB,CD}{}^{EF}&=&2\,\delta^{[E}_{[A}\theta_{B][C}\delta^{F]}_{D]}\,.
\end{eqnarray}
These are the well known ${\rm CSO}(p,q,r)$ gaugings originally
constructed in \cite{hull}, where $p,q,r$ ($p+q+r=8$) represent
the number of positive, negative and null eigenvalues of
$\theta_{AB}$ . For $p=8,q=0,r=0$ we have the ${\rm SO}(8)$
gauging constructed by de Wit and Nicolai \cite{dwn}. This gauging
is realized by setting $\theta_{AB}=\delta_{AB}$ namely
$\theta_{IJ}=\delta_{IJ}$, $\tau_I=0$, $\tilde{g}=1$. In general
the group ${\rm CSO}(p,q,r)$ describes the isometries of the
following seven dimensional hypersurface embedded in
$\mathbb{R}^8$ \cite{hw,hg,kallosh}:
\begin{eqnarray}
\theta_{AB}\,z^A\,z^B&=&R^2\,.
\end{eqnarray}
and which can be written locally as the product ${\Scr
H}^{p,q}\times \mathbb{R}^r$, ${\Scr H}^{p,q}$ being a
$(p+q)$--dimensional hyperboloid. The Maurer-Cartan equations for
this manifold have the form:
\begin{eqnarray}
dV^I+\omega^{IJ}\,\theta_{JK}\wedge
V^K-\tau_J\,V^J\wedge V^I&=&0\,,\\
d\omega^{IJ}+\omega^{IK}\theta_{KL}\wedge
\omega^{LJ}+2\,\omega^{K[I}\,\tau_K\,\wedge
V^{J]}-\tilde{g}\,V^I\wedge V^J&=&0\,,
\end{eqnarray}
$V^I$ and $\omega^{IJ}$ being the vielbein and the connection of
the manifold.\par  Together with the components (\ref{comp}) we
may switch on also the flux $g_{IJKL}$ in the $\overline{{\bf
35}}_{+5}$. The connection of the gauge algebra now becomes:
\begin{eqnarray}
\Omega_\mu
&=&A^I_\mu\,X_I+\tilde{A}^{IJ}_\mu\,X_{IJ}+A_{IJ\mu}\,X^{IJ}\,.
\end{eqnarray}
Let us consider the following basis of $\rE_{7(7)}$ generators:
\begin{eqnarray}
\{t_M{}^N,\,t^{MNP},\,t_{MNP},\,t^P,\,t_P\}\,.\label{basis}
\end{eqnarray}
We refer the reader to the appendix for the complete commutation
relations among the above generators.  With respect to the basis
(\ref{basis}) the gauge generators have the following expression:
\begin{eqnarray}
X_{MN}&=&-\theta_{[M|P}\,t_{N]}{}^P-\tau_{[M}\,t_{N]}\,,\nonumber\\
X_M&=&\frac{1}{2}\,(\tilde{g}\,t_M+\tau_N\,t_M{}^N+\theta_{MN}\,t^N)+2\,g_{MNPQ}\,t^{NPQ}+\frac{1}{14}\,\,\tau_M\,t\,,\nonumber\\
X^{MN}&=& 3\,g_{PQRS}\,\epsilon^{PQRSMNT}\,t_T\,.\label{gens}
\end{eqnarray}
where we have denoted by $t$ the ${\rm O}(1,1)$ generator
$t_M{}^M$. For this gauging the second order condition (\ref{2c})
read:
\begin{eqnarray}
\theta_{M[N}\,g_{PQRS]}&=&0\,\,\,;\,\,\,\,\tau_{[N}\,g_{PQRS]}=0\,,\label{2c2}
\end{eqnarray}
which imply that the $X^{MN}$ generators satisfy the following
constraints:
\begin{eqnarray}
\theta_{MN}X^{NP}&=&0\,\,\,;\,\,\,\,\tau_{N}\,X^{NP}=0\,.
\end{eqnarray}
Conditions (\ref{2c2}) guarantee that the generators  in
(\ref{gens}) close an algebra, which can be found to have the
following structure:
\begin{eqnarray}
\left[X_{MN},\,X_{PQ}\right]&=&\theta_{M[P}\,X_{Q]N}-\theta_{N[P}\,X_{Q]M}\,,\nonumber\\
\left[X_{MN},\,X_{P}\right]&=&\theta_{P[N}\,X_{M]}-\tau_{[N}\,X_{M]P}\,,\nonumber\\
\left[X_{M},\,X_{P}\right]&=&\tau_{[P}\,X_{M]}-\frac{1}{2}\,\tilde{g}\,X_{MP}+g_{MPNQ}\,X^{NQ}\,,\nonumber\\
\left[X_{M},\,X^{NP}\right]&=&3\,g_{P_1P_2P_3P_4}\,\epsilon^{NPP_1P_2P_3P_4S}\,X_{SM}-\frac{3}{14}\,\tau_M\,X^{NP}\,.
\end{eqnarray}
\paragraph{Third example: Scherk--Schwarz gauging}
As a second example let us consider the model originally studied
by Cremmer, Scherk and Schwarz \cite{css}, which describes a
generalized dimensional reduction of maximal $D=5$ supergravity to
$D=4$, in which the Scherk--Schwarz twist is chosen in
$\rE_{6(6)}$, global symmetry group of the five dimensional
theory. The resulting gauged supergravity \cite{adfl} is defined
by an embedding tensor transforming in the ${\bf 78}_{+3}$
\cite{dwst1} with respect to the $\rE_{6(6)}\times {\rm O}(1,1) $
subgroup of $\rE_{7(7)}$. In the basis of the ${\bf 56}$ in which
the 28 electric vector fields are
$A^\Lambda_\mu=\{A^u_\mu,\,A^0_\mu\}$, where $A^u_\mu$,
$u=1,\dots, 27$ are the dimensionally reduced five-- dimensional
vectors in the ${\bf 27}_{-1}$ of $\rE_{6(6)}\times {\rm O}(1,1) $
and $A^0_\mu$ is the Kaluza--Klein vector in the ${\bf 1}_{-3}$ of
the same group, the embedding tensor has just electric components
$\theta_\Lambda^\sigma$ and the gauge generators read:
\begin{eqnarray}
X_\Lambda &=&\cases{X_0=\theta_{0,u}{}^v\,t_{v}{}^u\cr
X_u=\theta_u{}^v\,t_v}\,,\nonumber\\
\theta_{0,u}{}^v&=&\theta_u{}^v=M_u{}^v\in \rE_{6(6)}\,.
\end{eqnarray}
where $M_u{}^v$ is the twist matrix depending in general on $78$
parameters, $t_u{}^v$ are the $\rE_{6(6)}$ generators, and $t_u$
are $\rE_{7(7)}$ generators in the $\overline{{\bf 27}}_{+2}$,
according to the following branching of the $\rE_{7(7)}$
generators with respect to $\rE_{6(6)}\times {\rm O}(1,1) $:
\begin{eqnarray}
{\bf 133}&\rightarrow & {\bf 78}_0+{\bf 1}_0+\overline{{\bf
27}}_{+2} +{\bf 27}_{-2}\,.
\end{eqnarray}
In this case the relevant components of the gauge generators are:
\begin{eqnarray}
X_{0u}{}^v&=&-X_{u0}{}^v=-X_{0}{}^v{}_u=X_u{}^v{}_0=-M_u{}^v\,,\nonumber\\
X_{uvw}&=&M_u{}^z\,d_{zvw}\,,\label{Xss}
\end{eqnarray}
where $d_{uvw}$ denotes the three times symmetric invariant tensor
of the ${\bf 27}$ of $\rE_{6(6)}$. To obtain eqs. (\ref{Xss}) we
have used the properties
$(t_u{}^v)^w{}_z=-(t_u{}^v)_z{}^w=\delta^w_u\delta^v_z-(1/27)\,\delta^w_z\delta^v_u$,
$(t_v)^w{}_0=-(t_v)_0{}^w{}=\delta^w_v$ and $(t_u)_{vw}=d_{uvw}$.
The gauge algebra has the following structure:
\begin{eqnarray}
[X_0,\,X_u]&=& M_u{}^v\,X_v\,,
\end{eqnarray}
all other commutators vanishing. \par If $M $ is non--compact the
corresponding theory depends effectively only on six parameters
and the potential is of run--away type, namely there is no vacuum
solution. If on the other hand $M$ is compact, the theory has
Minkowski vacua and depends effectively on four mass parameters,
which determine the amount of residual supersymmetry on these
solutions.
\section{Acknowledgements}
One of us (M.T.) would like to thank the UCLA Physics Department,
where the present investigation was completed, for its kind
ospitality and support. Work supported in part by the European
Community's Human Potential Program under contract
MRTN-CT-2004-005104 `Constituents, fundamental forces and
symmetries of the universe', in which R. D'A. and M.T.  are
associated to Torino University. The work of S.F. has been
supported in part by European Community's Human Potential Program
under contract MRTN-CT-2004-005104 `Constituents, fundamental
forces and symmetries of the universe', in association with INFN
Frascati National Laboratories and by D.O.E. grant
DE-FG03-91ER40662, Task C.
\appendix
\section{The $\rE_{7(7)}$ generators in the ${\rm
GL}(7,\,\mathbb{R})$ basis} We give below the commutation
relations among the $\rE_{7(7)}$ generators in the basis
(\ref{basis}):
\begin{eqnarray}
\left[t_M{}^N,\,t_P{}^Q\right]&=&\delta_{P}^N\,t_{M}{}^Q-\delta_{M}^Q\,t_{P}{}^N\,,\nonumber\\
\left[t_M{}^N,\,t^{P_1P_2P_3}\right]&=&-3\,\delta_M^{[P_1}\,t^{P_2P_3]N}+\frac{5}{7}\,\delta_{M}^N\,t^{P_1P_2P_3}\,,\nonumber\\
\left[t_M{}^N,\,t_{P}\right]&=&\delta_P^{N}\,t_{M}+\frac{3}{7}\,\delta_{M}^N\,t_{P}\,,\nonumber\\
\left[t^{N_1N_2N_3},\,t^{P_1P_2P_3}\right]&=&\epsilon^{N_1N_2N_3P_1P_2P_3
Q }\,t_Q\,,\nonumber\\
\left[t_M{}^N,\,t_{P_1P_2P_3}\right]&=&3\,\delta_{[P_1}^{N}\,t_{P_2P_3]M}-\frac{5}{7}\,\delta_{M}^N\,t_{P_1P_2P_3}\,,\nonumber\\
\left[t_M{}^N,\,t^{P}\right]&=&-\delta_M^{P}\,t^{N}-\frac{3}{7}\,\delta_{M}^N\,t^{P}\,,\nonumber\\
\left[t_{N_1N_2N_3},\,t_{P_1P_2P_3}\right]&=&\epsilon_{N_1N_2N_3P_1P_2P_3
Q }\,t^Q\,,\nonumber\\
\left[t^N,\,t_{M}\right]&=&t_M{}^N+\frac{1}{7}\,\delta_M^N\,t\,,\nonumber\\
\left[t^M,\,t^{N_1N_2N_3}\right]&=&-\frac{1}{6}\,\epsilon^{M
N_1N_2N_3 P_1P_2P_3}\,t_{P_1P_2P_3}\,,\nonumber\\
\left[t_M,\,t_{N_1N_2N_3}\right]&=&-\frac{1}{6}\,\epsilon_{M
N_1N_2N_3 P_1P_2P_3}\,t^{P_1P_2P_3}\,\nonumber\\
\left[t_{M_1M_2M_3},\,t^{N_1N_2N_3}\right]&=&18\,\delta^{[N_1N_2}_{[M_1M_2}\,t_{M_3]}{}^{N_3]}-\frac{24}{7}\,\delta^{N_1N_2
N_3 }_{M_1M_2 M_3}\,t\,,
\end{eqnarray}
where $t\equiv t_M{}^M$.

\end{document}